**A Molecule-Based Single-Photon Source Applied in Quantum Radiometry**

*Pietro Lombardi, Marco Trapuzzano, Maja Colautti, Giancarlo Margheri, Marco López, Stefan Kück, Costanza Toninelli\**

Dr. P. Lombardi,

Istituto Nazionale di Ottica (CNR-INO), via N. Carrara 1, Sesto F.no (FI) 50019, Italy

M. Trapuzzano,

Università degli Studi di Firenze, via G. Sansone 1, 50019 Sesto F.no (FI), Italy

M. Colautti,

LENS, Università degli Studi di Firenze, via N. Carrara 1, 50019 Sesto F.no (FI), Italy

Dr. G. Margheri,

Istituto dei Sistemi Complessi (CNR-ISC), via Madonna del Piano 10, 50019 Sesto F.no, Italy

Dr. M. López, Dr. S. Kück,

Physikalisch-Technische Bundesanstalt (PTB), Bundesallee 100, 38116 Braunschweig, Germany

Dr. C. Toninelli,

Istituto Nazionale di Ottica (CNR-INO), via N. Carrara 1, Sesto F.no (FI) 50019, Italy

E-mail: toninelli@lens.unifi.it

Single photon sources (SPSs) based on quantum emitters hold promise in quantum radiometry as metrology standard for photon fluxes at the low light level. Ideally this requires control over the photon flux in a wide dynamic range, sub-Poissonian photon statistics and narrow-band emission spectrum. In this work, a monochromatic single-photon source based on an organic dye molecule is presented, whose photon flux is traceably measured to be adjustable between 144 000 and 1320 000 photons per second at a wavelength of $(785.6 \pm 0.1)$ nm, corresponding to an optical radiant flux between 36.5 fW and 334 fW. The high purity of the single-photon stream is verified, with a second-order autocorrelation function at zero time delay below 0.1 throughout the whole range. Featuring an appropriate combination of emission properties, the molecular SPS shows here application in the calibration of a silicon Single-Photon Avalanche Detector (SPAD) against a low-noise analog silicon photodiode traceable to the primary standard for optical radiant flux (i.e. the cryogenic radiometer). Due to the narrow bandwidth of the source, corrections to the SPAD detection efficiency arising from the spectral power distribution are negligible. With this major advantage, the developed device may finally realize a low-photon-flux standard source for quantum radiometry.

**1. Introduction**

Sources of single photons are required for fundamental quantum optics experiments and are also key components in photonic quantum technologies.[1] Applications can be found in quantum cryptography,[2] quantum imaging,[3] simulation[4] and quantum-enhanced optical measurements.[5] Notably, they turn out to be ideal sources for radiometry, especially in quantum radiometry, where low photon fluxes (in the fW-range) have to be measured with low uncertainty. Indeed, current standards do not provide constant adjustable fluxes for calibrating single photon detectors and consequently all optical elements.[6] Even in the intermediate step of bridging the radiant power from single photon streams to the regime accessible with standard silicon photodiodes (calibrated against the primary standards), the problem of fluctuations in the average photon number per unit time arises in the case of attenuated lasers[7,8] and a sub-Poissonian photon stream would be desirable.

In principle, single-photon sources offer the possibility to realize a new primary standard for light sources[9] in the low-flux range, complementing the blackbody radiator and the synchrotron radiation source. It is conceptually simple to relate the photon flux $n$ with the energy flux, i.e. the optical radiant flux (optical power) $\Phi$ through the simple expression $\Phi = n\, h\nu = n\, hc/\lambda$, with $h$ the Planck constant, $c$ the speed of light, $\nu$ the frequency and $\lambda$ the wavelength of the emitted radiation. In case of pulsed excitation and ideal photon source, $n$ is exactly the pump repetition rate $f$, whereas for continuous wave (CW) pumping, the maximum photon flux is essentially approximately determined by the inverse of the excited state lifetime ($1/\tau$). Although in practice single-photon sources are never ideal due to the non-unitary collection and quantum efficiency, they may provide a reproducible photon flux once metrologically characterized. In fact, from the quantum theory of photodetection (see e.g. Ref. [10]), considering the photon count statistics measured in a time interval of $\Delta t$, the relationship between the variance in the measured photon flux ($\Delta N^2$) and the corresponding variance in the emitted photon flux ($\Delta n^2$) in the same time interval is given by

$$(\Delta N)^2 = \eta^2 (\Delta n)^2 + \eta(1-\eta) <n>/\Delta t \,, \tag{1}$$

where $\eta$ is the overall emission, collection and detection efficiency. In other words, the advantage of using photon Fock states with respect to weak coherent pulses, when $<n> \Delta t = 1$, is accounted for by the factor $(1-\eta)$ in the variance of the measured photon flux.

First experiments in this direction have been performed with color centers in diamond. The source presented in Ref. [11], e.g., is based on an NV-center doped nano-diamond at room temperature, and is hence not particularly suitable for application in single-photon detector calibration, because of the broad emission spectrum, ($\Delta\lambda_{FWHM} \sim 100\ nm$). A much narrower bandwidth source is reported in Ref. [12], based on a silicon-vacancy center. In this case, however, a direct calibration could not be carried out since the low photon rate (60 kphotons/s), due to the very small $\eta$. In this respect, sources based on semiconductor quantum dots[13] or molecules[14] in cryogenic environment look currently more suitable for applications. In particular, polycyclic aromatic hydrocarbon molecules show an unmatched combination of suitable properties: quantum yield close to unity, pronounced branching ratio in favor of the narrowband zero phonon line (00-ZPL) at cryogenic temperature and photostable emission.[15]

In this paper, we report on the optimized and metrologically characterized photon flux from a single-photon source based on a dibenzoterrylene (DBT) molecule in an anthracene nanocrystal, exhibiting strong sub-Poissonian photon statistics, as well as narrow-band, bright and photostable emission. The source is hence deployed under CW excitation for the calibration of a single-photon detector directly against a classical silicon photodiode, which is in turn traced to the primary standard for optical radiant flux, i.e. the cryogenic radiometer.

In Section 2, the molecule-based single-photon source used for the experiments is described. Section 3 deals with the metrological characterization of the source, while in Section 4 the

SPS-based calibration of a silicon SPAD is discussed. In Section 5 we draw conclusions and outlooks, and in Section 6 the experimental steps are described in more detail.

## 2. The single-photon source

DBT molecules in anthracene (Ac) emit narrow-band photons when cooled down to cryogenic temperatures, exhibiting high quantum efficiency, photostability and quantum coherence, [16,17] even embedded in small nanocrystals.[18] However, since a high photon flux is relevant for applications in quantum radiometry, collection efficiency of single molecule fluorescence requires optimization. This is particularly true for low numerical aperture optics, such is the case of the long-working-distance microscope objective (N.A.=0.67), which is used in the setup for cryogenic operation. In order to maximize the detected photon flux from single DBT molecules in Ac nanocrystals at low temperatures, the multilayer configuration discussed in Ref. [19] is further adapted.

The single-photon source used in this work is obtained from an isolated DBT molecule, placed around 100 nm away from a metallic mirror (160-nm thick gold layer). This distance fits within the $\lambda/(6n) \leftrightarrow \lambda/(4n)$ interval (where $\lambda$ is the emission wavelength and $n$ the refractive index of the medium), i.e. the condition which maximizes the emission directionality within a small angle around the polar axis.[20] This simple configuration can be obtained with a cost-effective procedure, based on the deposition of DBT-doped Ac nano-crystals over the gold mirror. Such nano-crystals are prepared as suspension in water through a reprecipitation method,[18] which enables a certain control over the crystal size and emitter concentration. For the device under investigation, we grow 200-nm thick nanocrystals containing single DBT molecules. The fabrication protocol is terminated by spin-coating a 200-nm thick PVA layer, in order to stabilise the sample and flatten the interface between the dielectric layer and air.

Fine adjustment of the molecule distance with respect to the mirror is obtained exploiting their almost flat statistical distribution of the position inside the host matrix. Indeed, the sample provide millions of single-photon sources, out of which few thousands appear brighter, due to their optimal placement and orientation with respect to the metallic mirror. Thanks to a very low probability of having more than one molecule per nanocrystal, such relevant cases correspond to the brightest spots of a fluorescent map such as the one shown in Figure 1b, obtained under wide field illumination and imaging on camera.

In Figure 1c, a sketch of the optical setup is outlined, representing an epifluorescence microscope, where the sample is cooled down to 3 K and which is equipped with different detection options. More details about the optical setup are provided in Section 6.

Once a promising nanocrystal is selected, confocal illumination (and detection) is adopted for the optical characterization of the source. The use of isolated nano-crystals enables single emitter addressing without specific spatial filtering beyond confocal microscopy.

## 3. Metrological characterization of the molecule-based single-photon source

Single molecules are pumped into the first electronic excited stated (which has a lifetime $\tau \sim 4$ ns) via an auxiliary vibrational level using a diode laser centered at a wavelength of 767 nm (see Figure 1a). The Stoke-shifted fluorescence is filtered out and characterized in terms of the photon statistics and spectral features. In Figure 2a the emission spectrum of the molecule which is metrologically characterized and deployed in this paper is shown. The peak around 767 nm is due to the residual laser light, whereas the most intense signal is associated to the molecule main transition, i.e. its 00-ZPL. This is filtered in a bandwidth of about 2 nm around the wavelength $\lambda \sim 785$ nm. In the inset of Figure 2a the resulting spectrum appears limited by the spectrometer resolution (~ 0.2 nm) and can be independently measured via excitation spectroscopy to be smaller than one picometer.[18] The measurements reported in the following sections are obtained in these operative conditions.

Both photon flux and $g^{(2)}(t)$ function are measured at the output of a multimode fiber, as a function of the excitation power. Extract from the measurement results are depicted in Figure 2b and 2c-d, respectively, for the molecule chosen as a reference.

The molecular-based SPS is able to deliver at the fiber-coupled detector up to $1.4 \times 10^6$ photons per second, keeping high purity of the single-photon emission for any set rate (in particular, $g^{(2)}(0) = 0.08 \pm 0.01$ at maximum photon rate, without deconvolving for the SPAD response function of around 0.4 ns). These characteristics are outstanding especially when considering the detected power in a given frequency interval. Indeed, according to excitation spectroscopy reported in our previous works,[18] around 2/3 of the collected photon flux, i.e. more than $0.9 \times 10^6$ photons/s, falls within a 50MHz-wide spectral window. Overall the developed source fulfils the requirements to operate as a secondary standard source for SPAD calibration (see Section 4).

It should be noted that the saturation curve reported in Figure 2b decays for very high pump powers. This power-dependent shelving effect has not been reported before for organic molecules and is currently under investigation.[21] For the purpose of this paper however, it only limits the accessible photon flux, which is already relevant for radiometric applications.

The photon statistics of the source is obtained measuring the histogram of the difference in the photon arrival times using start and stop signals from two SPAD detectors, arranged in a Hanbury-Brown Twiss configuration (see Figure 5). It is well known that such data set represents correctly the $g^{(2)}(t)$ only for short times. Indeed, for long times the coincidence probability is suppressed by the high detected count rate.

The $g^{(2)}(t)$ function can be fitted with the expression

$$g^{(2)}(t) = \left(1 - b * e^{-|t|/t_1}\right) * e^{-R*t} \qquad (2)$$

where $b$ and $t_1$ represent the depth and the time constant of the anti-bunching dip, respectively,[22] and the last term accounts for the effect of the arrival time statistics, considering Poissonian distribution ($R$ is the average count rate per SPAD). We can exclude in first approximation a bunching contribution given by the dark periods related to the occupation of the meta-stable triplet state (blinking), since it holds on several microsecond time scale, and off-time is estimated to be limited to few percent.[16] Indeed, the value for $R$ obtained by the fit is in agreement with the count rate directly read by each detector.

In the framework of metrological applications, it is relevant to determine the stability of the photon flux under CW excitation. We report less than 1% fluctuations on short time-scale (seconds), and a drift of around 2% due to the pump light alignment within a time interval of 10 minutes.

Another set of measurements has been devoted to the determination of the highest temperature at which the device is able to guarantee a photon stream with reasonable optical properties for metrological applications, such as: photon flux > 1 Mcps; $g^{(2)}(0) < 0.1$; spectrum FWHM < 2 nm. [23]

Raising the temperature indeed, line-broadening is expected, together with a lowering of both the absorption cross section and of the branching ratio into the 00-ZPL. In Figure 3, the emission spectrum for different temperatures is shown for a fixed pump power. Due to the limited resolution of the spectrometer (evaluated to be around 0.2 nm by measuring a laser line with actual linewidth < 10 fm), broadening of the line becomes evident only around 20 K. However, the combined effect of the other two aspects is effective already at 10 K and can be only partially mitigated by a stronger excitation. Table 1 gathers the results of this analysis, which fixes the maximal operating temperature of the device to approximatively 20 K.

## 4. Calibration of a single-photon detector with a molecule-based SPS

The detection efficiency of a Si-SPAD detector is determined by comparing the photon flux measurements of the single photon source (see Section 2) performed with the SPAD detector (device under test) and an analog reference Si-detector. Details about the traceability chain for the calibration of the latter device are given in Section 6. The reference detector and the SPAD are equipped with a FC/PC fiber connector and their coupling efficiency is optimized for a multimode fiber. The photon flux measurements are performed sequentially. Thus, the SPAD detection efficiency $\eta_{SPAD}$ is determined as $\eta_{SPAD} = \frac{<N_{SPAD}>}{<N_{ref}>}$, where $N_{SPAD}$ is the count rate (counts/s) measured with the SPAD detector, while $N_{ref}$ is the photon flux rate derived from the source optical flux measurement $\Phi_s$ and the photon energy $E$ ($E = 2.53\times10^{-19}$ J for photon at 785.6 nm). $\Phi_s$ is obtained as the ratio between the measured average photocurrent $<I_f>$ and the reference detector responsivity $s_{ref}$, and hence $<N_{ref}> = \frac{<\Phi_s>}{E} = \frac{<I_f>/s_{ref}}{E}$.

Figure 4 shows the detection efficiency obtained for the SPAD detector (Perkin Elmer, SPCM-AQRH-13-FC) within the photon rate range from 0.144 Mphoton/s to 1.32 Mphoton/s, which corresponds to an optical power range between 36.5 fW to 334 fW. To the best of our knowledge, such broad flux interval was never explored so far with a single photon source. Interestingly, the molecule emission rate approaches the regime in which the detector dead time ($\tau_{dead}$) affects the measurement of the detection efficiency $\eta_{SPAD}$ [8].

The standard uncertainty associated with each measurement value is indicated by an error bar. It was calculated following the guidelines expressed in Ref.[24]. The achieved uncertainty varies within the range from 2 % to 6 %, depending on the photon rate, i.e. the lower the photon rate, the higher is the uncertainty. This can be ascribed to the reference detector random noise, which is the highest contribution to the total uncertainty at fW-levels, as observed in the uncertainty budget shown in Table 2. The final value obtained for the Si-SPAD quantum efficiency is $\eta_{SPAD} = (0.603 \pm 0.012)$.

## 5. Conclusions

In this paper we demonstrate the realization of an absolute single-photon source based on the emission of an organic dye molecule operated at cryogenic temperature. This result is obtained by linking the single-photon stream generated by the molecule to a national radiometric standard for optical fluxes via an analog Si-detector, calibrated through an unbroken traceability chain and able to read optical radiant fluxes down to a few tens of fW. The source presented here shows significant advances with respect to previous demonstrations in the field of radiometry in terms of flux (1.32 Mphoton/s), linewidth (< 0.2nm) and purity ($g^{(2)}(0) < 0.1$ ) of the emission. With this major advantage, the developed device may finally realize a low-photon-flux standard source for quantum radiometry, complementing the blackbody radiator and the synchrotron radiation source.

The traceably measured optical radiant flux adjustable between 37 fW and 334 fW at a wavelength of $(785.6 \pm 0.1)$ nm is unprecedented and allows the direct calibration of a single-photon detector (SPAD) through comparison with calibrated high-sensitivity analog Si detector for the first time.

The reported single-photon source can in principle be operated in pulsed conditions, with an estimated photon flux of around 300 kphoton/s for 20 MHz pump repetition rate. Under these conditions the device can works as a predictable true single-photon source whose photon flux is directly tuned by acting on the pump repetition rate, with high reliability and precision also at power levels below the detection limit of conventional photodetectors.

Beyond the demonstrated impact in quantum radiometry, thanks to the long coherence time and the efficient molecular emission into the Fourier-limited 00-ZPL, the presented device might find immediate applications also for quantum communication, simulation and computing, or in quantum imaging.

## 6. Experimental Section

*Optical setup*

The epifluorescence microscope setup used in the experiment is shown schematically in Figure 5. For CW excitation of the DBT molecules, an external cavity diode laser operating at a wavelength of 767 nm (Toptica DLX110) is employed. The laser beam is first spatially filtered (through coupling into a PM fiber, Thorlabs P3-780PM-FC) in order to fit a gaussian profile; then it is spectrally filtered (with the bandpass filter BP1, Semrock Bightline FF01-769/41) in order to avoid residual emission at the detection wavelength leaking into the detection path; finally it is mode-matched with the objective lens back entrance (SigmaKoki PAL-50-NIR-HR-LC07, N.A.= 0.67, transmission at 785 nm = 0.7) through the appropriate two-lens telescope, in order to exploit the available numerical aperture and minimize the size of the confocal spot (evaluated around 1 μm in diameter). A beam sampler BS (Thorlabs BSF20-B) with ~ 0.1/0.9 reflection/transmission coefficients is used to reflect the light towards the objective lens and conversely transmit the incoming signal. In the common path, a telecentric lens system linking a mirror with motorized tilt and the objective back entrance allows for exploration of the sample without the need for a translational stage. Another converging lens can be added before the BS if wide-field illumination is required ($L_{WF}$). The sample is fixed in thermal contact with the cold finger of a closed-loop liquid-helium cryostat (Montana Instrument) and kept at 3 K, and is optically addressable though a double window for a total glass thickness of 0.7 mm. The objective lens in use is designed to compensate for that.

The signal transmitted by the BS is directed towards the detection box, which is equipped at its entrance with a longpass filter (LP, Semrock RazorEdge LP02-785RE-25) to filter out the laser light back reflected by the sample. A second bandpass filter (BP2, Semrock TBP01-

790/12), whose transmission window can be shifted to the blue by tilting, is added in the detection path if selection of the 00-ZPL line is required. The photon flux is hence redirected to a fiber coupling system consisting in a low magnification objective lens (Olympus UPlanFL N, 10x, N.A.= 0.30, transmission at 785 nm = 0.8) focusing light into a cleaved multi-mode fiber (Thorlabs M42L02, core diameter: 50 µm, N.A.= 0.22). The fiber can deliver the photon stream to either a low noise analog Si detector (Femto FWPR-20-s), or to a calibrated Si SPAD (Perkin Elmer SPCM-AQR-13-FC), or to a couple of SPAD in Hambury-Brown and Twiss (HBT) configuration (Excelitas 800-14-FC) through a fibered beam splitter (Thorlabs FCMM625-50A-FC). The latter arrangement, with the help of a time-correlated single-photon counting system (PicoQuant PicoHarp 300), gives access to the second-order autocorrelation function of the photon stream $g^{(2)}(t)$ in the form of time delay histogram of the detection events between the two detectors.

Finally, a flippable mirror set before the fiber coupling system can deviate the signal beam towards an EM-CCD-equipped spectrometer (Andor Shamrock SR-303i-A, camera iXon3), which can work also as simple wide-field imaging camera (tube-lens $L_T$ focal length 20cm). By adding a second converging lens $L_{BFP}$ at the appropriate position between the tube-lens and the camera, imaging of the back focal plane of the objective lens is obtained.

*Traceability chain for the calibration of the reference Si-detector*

The reference detector used for the photon flux measurement of the source is an analog ultra-low noise Si-detector (Femto FWPR-20-s). It consists of a fibre coupled Si photodiode of 1.1 x 1.1 mm$^2$ active area and a trans-impedance amplifier with gain of $1 \times 10^{12}$ V/A. The minimum noise equivalent power (NEP) of the detector is 0.7 fW/Hz$^{1/2}$. Its spectral responsivity $s_{Si}(\lambda)$ is determined by calibrating it against a working standard traceable to PTB's primary standard for optical power, i.e. the cryogenic radiometer.[25] The complete traceability chain is shown in Figure 6, where $\lambda_i$ is the wavelength, $\Phi$ is the optical radiant flux and $u$ is the standard measurement uncertainty. In a first step, a Si trap detector is calibrated against the cryogenic radiometer at specific wavelengths and optical powers (μW). Then, a spectrally flat detector, i.e. a thermopile detector, is used to determine the responsivity of a Si photodiode, which acts as working standard at the specific wavelength of the photons emitted by the single photon source, i.e. 785.6 nm. Finally, the spectral responsivity of the low-noise Si-detector used for the calibration of the Si-SPAD detector is obtained by means of the double attenuation calibration technique described in Ref.[8] and the Si photodiode working standard. The spectral responsivity obtained at 785.6 nm is,

$$s_{Si} = (57.52 \pm 0.58) \times 10^{-2} \ A/W \qquad (4)$$

The uncertainty of the spectral responsivity here reported corresponds to a standard uncertainty ($k = 1$).

**Acknowledgements**

The work reported on this paper was funded by the project EMPIR 17FUN06 SIQUST. This project received funding from the EMPIR program co-financed by the Participating States and from the European Union Horizon 2020 research and innovation program. The authors would like to thank Dr. Ivo Pietro De Giovanni for his advice and fruitful discussion on the role of photon statistics in the calibration of single-photon avalanche photodiodes and related measurements.

Received: ((will be filled in by the editorial staff))

Revised: ((will be filled in by the editorial staff))

Published online: ((will be filled in by the editorial staff))

Figures

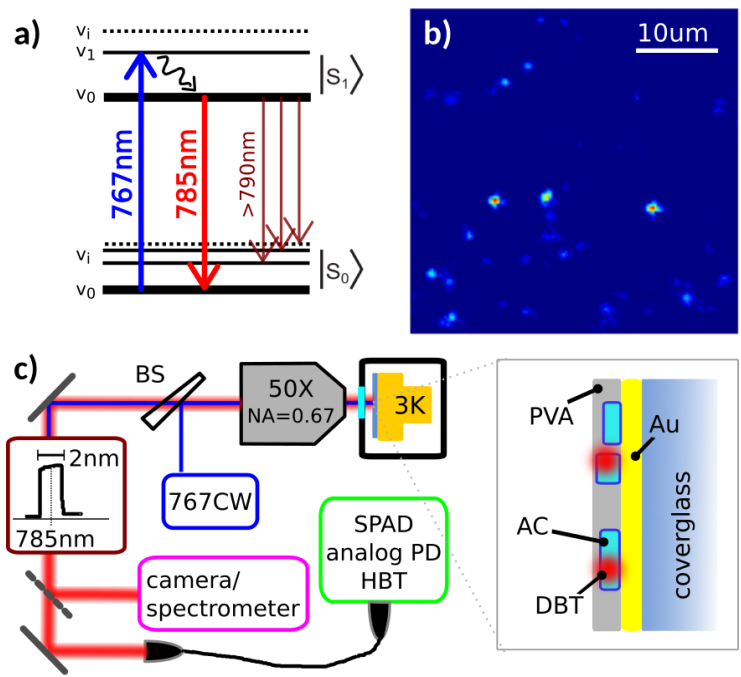

**Figure 1** a) Blue-detuned (767nm) pumping scheme used to collect photons emitted into the 00-ZPL (785nm). b) wide-field (WF) fluorescence image: zoom (40x40 um$^2$) on a region of the sample showing bright nanocrystals; c) simplified sketch of the optical setup used for the measurement reported in the paper (details are presented in Section 6) and sketch of the device operated as single-photon source: Au – gold, AC – anthracene, DBT – dibenzoterrylene, PVA – polyvinyl-alcohol.

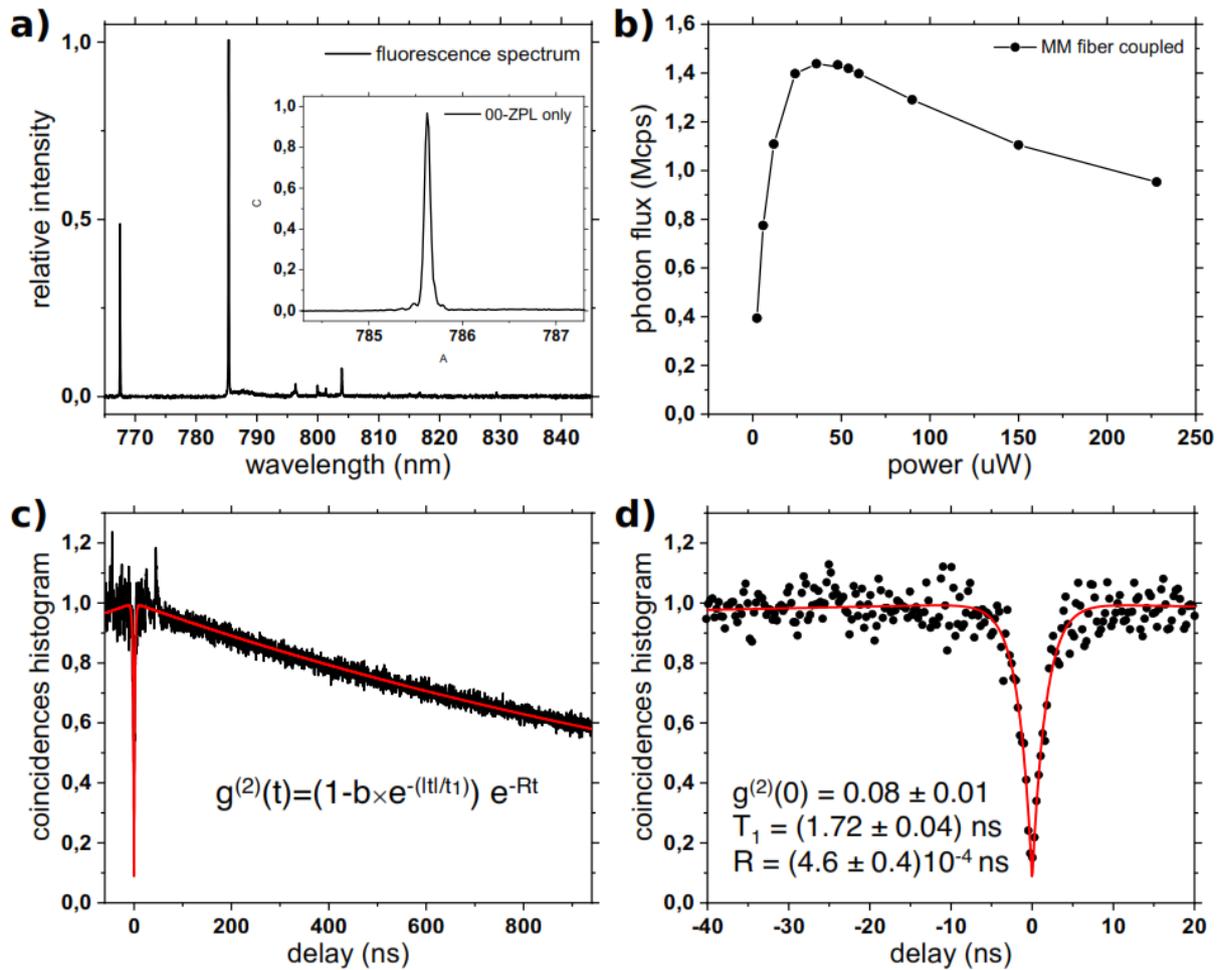

**Figure 2.** Metrological characterization of the molecule emission: a) Fluorescence spectrum, inset: fluorescence spectrum when filters are set to select a 2nm-wide spectral window around the molecule 00-ZPL (785.6nm in this case). b) Photon flux detected with the SPAD as a function of the laser pump power. c) Normalized histogram of the inter-photon arrival times for maximum photon flux operation (30 µW pump power). d) Zoom on the histogram in c) around zero time delay, representing $g^{(2)}(t)$: the anti-bunching behaviour shows the high purity of the single-photon stream. The red lines are a fit to the data with the expression shown in c), while best estimation of the fit parameters is reported in d).

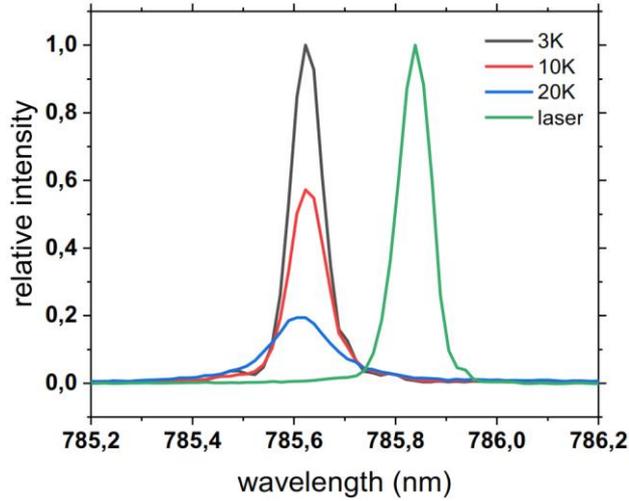

**Figure 3.** Spectral response of the narrow-band line as a function of temperature, for a fixed pump power equal to 30 µW. A lower absorption cross-section is observable already at 10 K, while spectral broadening is evident at higher temperature only, due to the limited spectrometer resolution (~ 0.2 nm estimation from laser line (green curve)).

**Table 1.** Attainable flux and purity of the single-photon stream for different temperatures

| Temperature [K] | Power [a)] [µW] | Max counts [Mphotons/s] | $g^{(2)}(0)$ |
|---|---|---|---|
| 3 | 30 | 1.36 | 0.08+/-0.01 |
| 5 | 42 | 1.27 | |
| 10 | 42 | 1.20 | |
| 15 | 42 | 1.09 | 0.06+/-0.02 |
| 15 | 72 | 1.19 | |
| 20 | 72 | 1.08 | 0.09+/-0.02 |

a) power measured at the entrance of the objective lens

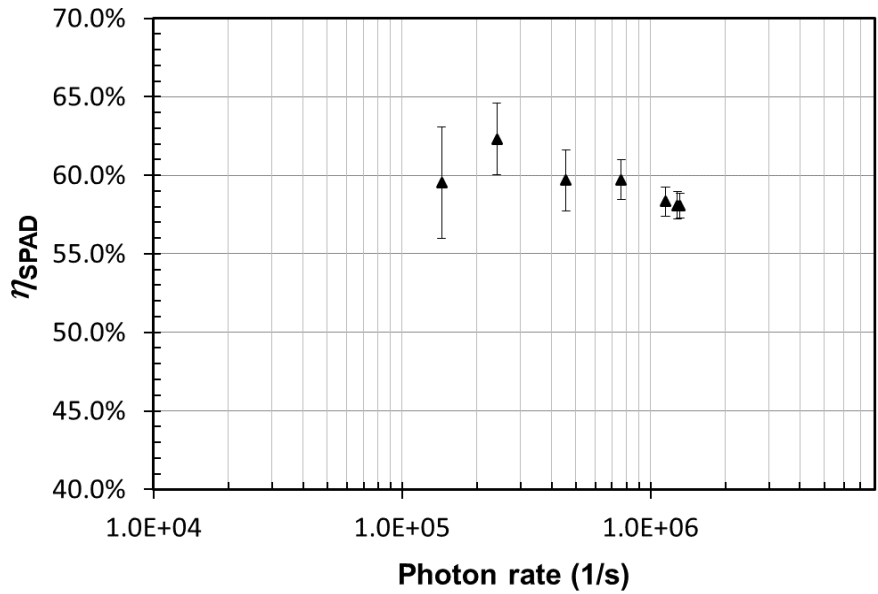

**Figure 4** Calibration result for the SPAD detection efficiency (Perkin Elmer, SPCM-AQRH-13-FC) using the molecule-based single-photon source and a low-noise reference analog detector.

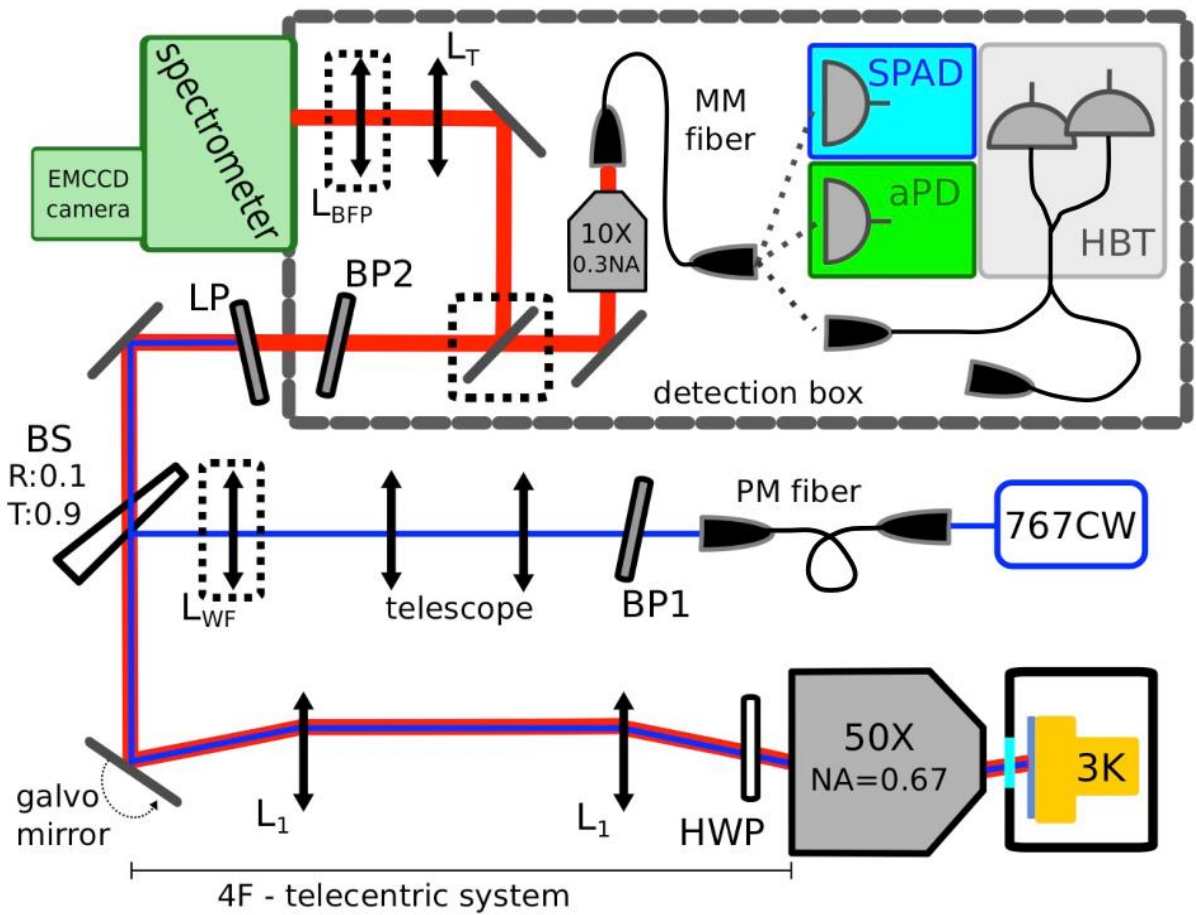

**Figure 5** Detailed sketch of the optical setup: dashed squares mark flippable elements; PM - polarization maintaining fiber, BP - bandpass filter, $L_{WF}$ - lens for wide-field imaging, BS - beam sampler, $L_1$ - lenses for telecentric system, HWP - half wave plate, LP - longpass filter, $L_T$ - tube lens, $L_{BFP}$ - lens for back focal plane imaging, MM - multi-mode fiber, SPAD - single photon counting module, aPD - analog Si photodiode, HBT - fiber-based Hanbury-Brown and Twiss interferometer.

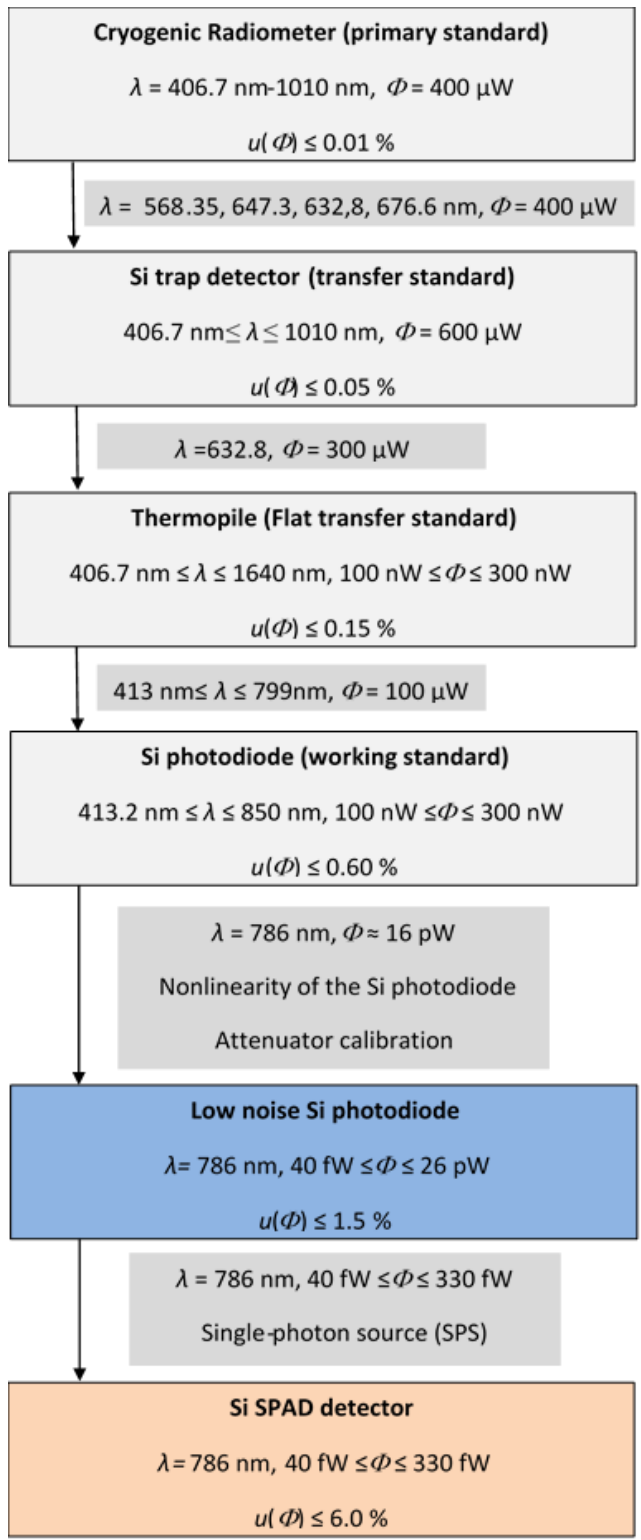

**Figure 6.** Traceability chain for the calibration of the low-noise analog Si-detector, and the additional link reported in this paper, which opens to the traceability of optical power flux down to few hundreds of aW.

**Table 2.** Uncertainty budget for the Si-SPAD detection efficiency $\eta_{SPAD}$ determined for an optical power of approx. 193 fW, which corresponds to a photon flux rate of approx. 764 kphoton/s. The model used for the estimation of the uncertainty is given by $\eta_{SPAD} = \frac{hc}{\lambda} \cdot \frac{S_{Si} F_{Amp} SPAD_{Counts}}{V_f (1-F_{Lin})}$, where $h$ is the plank´s constant, $c$ is the speed of the light, $\lambda$ is the wavelength, $F_{Amp}$ is the amplification factor of the internal amplifier of the reference detector, $V_f$ is the photo-voltage measurement (Si-detector measurement), $F_{Lin}$ is the linearity factor correction of the Si reference detector and $SPAD_{Counts}$ is the SPAD counts including dark counts correction.

| Source of uncertainty | Standard uncertainty (%) |
|---|---|
| Planck´s constant, $h$ | - |
| Wavelength, $\lambda$ | 0.008 |
| Speed of light, $c$ | - |
| Si-detector spectral responsivity, $s_{Si}$ | 0.400 |
| Si-detector measurement, $V_f$ | 1.870 |
| Amplification factor, $F_{Amp}$ | 0.100 |
| Linearity factor of the Si-detector, $F_{Lin}$ | 0.030 |
| Si-SPAD Counts, $SPAD_{Counts}$ | 0.020 |
| **Combined uncertainty, $u_c$** | **1.92** |

**Table of contents**

Here a molecule-based single photon source ready for application in quantum metrology is presented. The source is linked to a national radiometric standard for optical fluxes, assessing the device in terms of an absolute single-photon source. Finally it is deployed for the direct calibration of a single-photon avalanche detector, on a large range of optical radiant fluxes.

**Keyword:** quantum radiometry

*Pietro Lombardi , Marco Trapuzzano, Maja Colautti, Giancarlo Margheri, Marco López, Stefan Kück, Costanza Toninelli\**

**Title: A Molecule-Based Single-Photon Source Applied in Quantum Radiometry**

**ToC figure:**

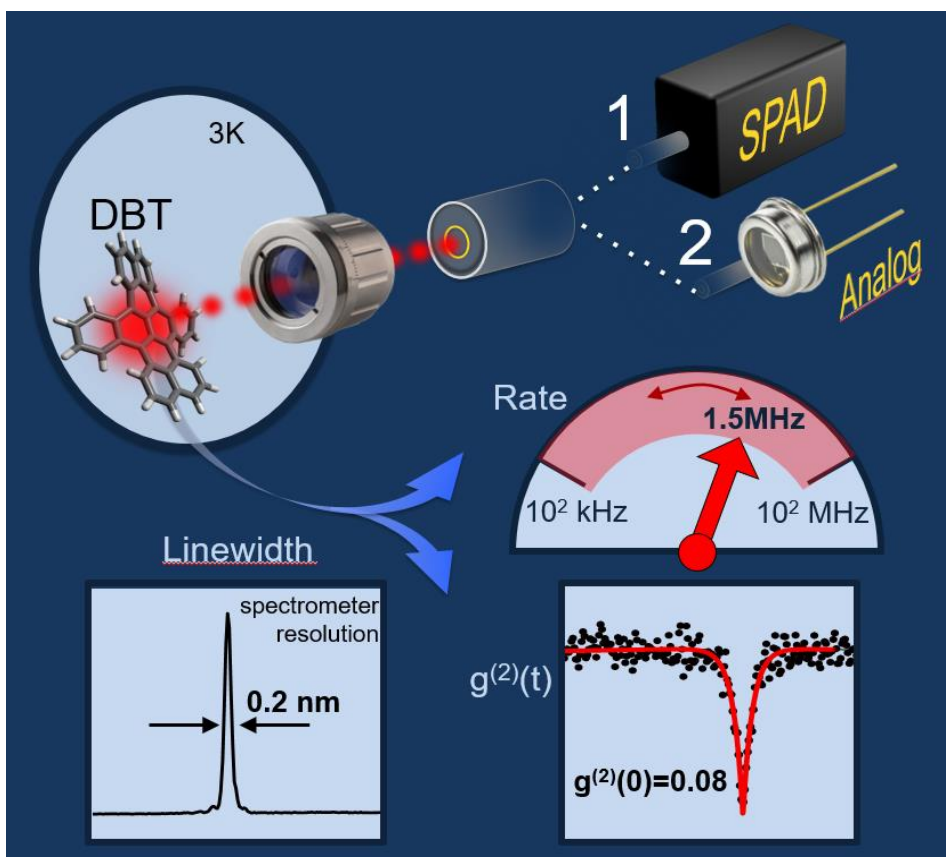